\pdfoutput=1

\documentclass[11pt,twoside,a4paper,cmspaper,final,collab]{cms-tdr}
\def\svnVersion{d8f7f79-D}\def\svnDate{2020/04/24}\def\cmsCernNoTag{CERN-EP-2019-279}\def\cmsCernDate{\today}\def\cmsMessage{"Published in Physical Review Letters as \href{http://dx.doi.org/10.1103/PhysRevLett.124.162002}{\doi{10.1103/PhysRevLett.124.162002}.}"}

\begin{document}\cmsNoteHeader{BPH-13-001}

\hyphenation{had-ron-i-za-tion}
\hyphenation{cal-or-i-me-ter}
\hyphenation{de-vices}

\newlength\cmsFigWidth
\ifthenelse{\boolean{cms@external}}{\setlength\cmsFigWidth{0.75\columnwidth}}{\setlength\cmsFigWidth{0.45\textwidth}}
\newlength\cmsFigWide
\ifthenelse{\boolean{cms@external}}{\setlength\cmsFigWide{\columnwidth}}{\setlength\cmsFigWide{0.7\textwidth}}
\newlength\cmsTabSkip\setlength\cmsTabSkip{10pt}

\ifthenelse{\boolean{cms@external}}{\providecommand{\suppMaterial}{the supplemental material~\cite{bib:suppMaterial}\xspace}} {\providecommand{\suppMaterial}{Appendix~\ref{app:tables}\xspace}}
\ifthenelse{\boolean{cms@external}}{\providecommand{\suppMaterialRef}[2]{#1}}{\providecommand{\suppMaterialRef}[2]{#2}}

\newcommand{\lth}{\ensuremath{\lambda_\vartheta}\xspace}
\newcommand{\lph}{\ensuremath{\lambda_\varphi}\xspace}
\newcommand{\ltpp}{\ensuremath{\lambda_{\vartheta\varphi}}\xspace}
\newcommand{\abscosth}{\ensuremath{\abs{\cos\vartheta}}\xspace}
\newcommand{\Rcosth}{\ensuremath{R(\abscosth)}\xspace}
\newcommand{\Rphi}{\ensuremath{R(\varphi)}\xspace}

\newcommand{\Chic}{\ensuremath{\chi_{\PQc}}\xspace}
\newcommand{\Chib}{\ensuremath{\chi_{\PQb}}\xspace}
\newcommand{\chic}[1]{\ensuremath{\chi_{{\PQc}#1}}\xspace}
\newcommand{\ChicZero}{\ensuremath{\chic{0}}\xspace}
\newcommand{\ChicOne}{\ensuremath{\chic{1}}\xspace}
\newcommand{\ChicTwo}{\ensuremath{\chic{2}}\xspace}
\newcommand{\ChicJ}{\ensuremath{\chic{J}}\xspace}

\providecommand{\PQQ}{\HepParticle{Q}{}{}\Xspace} 
\providecommand{\PAQQ}{\HepAntiParticle{Q}{}{}\Xspace} 
\providecommand{\QQbar}{\ensuremath{\PQQ\PAQQ}\xspace}

\title{\texorpdfstring
{Constraints on the \ChicOne versus \ChicTwo polarizations
in proton-proton collisions at $\sqrt{s} = 8$\TeV}
{Constraints on the chic1 versus chic2 polarizations
in proton-proton collisions at sqrts = 8 TeV}}

\date{\today}

\abstract{
The polarizations of promptly produced \ChicOne and \ChicTwo mesons are studied
using data collected by the CMS experiment at the LHC, in proton-proton collisions at $\sqrt{s} = 8$\,TeV.
The \Chic states are reconstructed via their radiative decays $\Chic \rightarrow \JPsi\, \gamma$,
with the photons being measured through conversions to $\Pep\Pem$,
which allows the two states to be well resolved.
The polarizations are measured in the helicity frame,
through the analysis of the \ChicTwo to \ChicOne yield ratio
as a function of the polar or azimuthal angle of the positive muon emitted in the $\JPsi \to \mu^+\mu^-$ decay,
in three bins of \JPsi transverse momentum.
While no differences are seen between the two states in terms of azimuthal decay
angle distributions, they are observed to have significantly different polar anisotropies.
The measurement favors a scenario where at least one of the two states is strongly polarized
along the helicity quantization axis, in agreement with nonrelativistic quantum chromodynamics predictions.
This is the first measurement of significantly polarized quarkonia produced at high transverse momentum.}

\hypersetup{%
pdfauthor={CMS Collaboration},%
pdftitle={Measurement of the chic1 and chic2 polarizations in proton-proton collisions at sqrt(s) = 8 TeV},%
pdfsubject={CMS},%
pdfkeywords={CMS, physics, quarkonium polarization}}

\maketitle

Quarkonium production is a benchmark for understanding how quarks combine into hadrons.
The heaviness of \cPqc and \cPqb quarks makes it possible to describe the process in
nonrelativistic quantum chromodynamics
(NRQCD)~\cite{bib:NRQCD,Butenschoen:2012px,Butenschoen:2012qr,Chao:2012iv,Gong:2012ug,Shao:2014fca,Shao:2012fs,Bodwin:2015iua},
a framework valid when the quark velocities are small.
This theory successfully described quarkonium cross sections measured~\cite{Abe:1997jz}
at high transverse momentum, \pt, by complementing the
earlier color-singlet model~\cite{Baier:1981uk,bib:lansberg-HP08}
with a superposition of several processes where the bound state originates from colored \QQbar pairs.
In contrast to this complex model,
the \JPsi, $\Pgy$, $\PgUa$, $\PgUb$, and $\PgUc$
differential cross sections measured at central rapidity by
ATLAS~\cite{bib:ATLASpsi2S,bib:ATLASYnS} and CMS~\cite{bib:BPH-14-001,bib:BPH-12-006,bib:BPH-15-005}
have indistinguishable shapes as a function of $\pt/M$, where $M$ is the meson mass~\cite{bib:Faccioli:PLB773,bib:Faccioli:EPJC78p118}.
This universal momentum scaling pattern is also followed by the \ChicOne and \ChicTwo states~\cite{bib:ATLASchic,bib:BPH-11-010}.
The corresponding polarization measurements~\cite{bib:BPH-13-003,bib:BPH-11-023} show that
the five S-wave
states are well compatible with being produced unpolarized,
in contrast to the significant polarizations seen for the \PW and
\PZ~\cite{Mirkes:1994eb,Aaltonen:2011nr,Khachatryan:2015paa,Abbott:2000aj,Acosta:2003jm,Acosta:2005dn,Chatrchyan:2011ig,ATLAS:2012au},
Drell--Yan dileptons~\cite{Lam:1978pu,Faccioli:2010ej,Guanziroli:1987rp,Conway:1989fs,Zhu:2006gx,Zhu:2008sj},
and low-\pt quarkonia~\cite{Abt:2009nu,Brown:2000bz}.
The lack of polarization of high-\pt vector quarkonia was a long-standing challenge for NRQCD~\cite{Brambilla:2010cs},
until recent global-fit analyses~\cite{Chao:2012iv,bib:Faccioli:PLB736,Bodwin:2014gia}
showed that cross sections and polarizations can be consistently described,
unveiling a delicate compensation between terms in the factorization expansion~\cite{bib:Faccioli:2019hht}.
Among the measurements mentioned above, one piece is clearly missing:
the \ChicOne and \ChicTwo polarizations.
Contrary to what happens for the vector states,
predicting the \ChicOne and \ChicTwo polarizations is rather simple within NRQCD,
where they are unequivocally determined by a single color-octet parameter,
which can be extracted from the \ChicTwo to \ChicOne cross section ratio.
The analysis of the measured ratios~\cite{bib:ATLASchic,bib:BPH-11-010} provides a clear result:
the polarizations of the two states should be opposite and almost maximal~\cite{bib:Faccioli:EPJC78p268}
(a result also reached in a parameter-free singlet-only model~\cite{Baranov:2019orn}).
Finding that these P-wave states have similar polarizations
(following the vector quarkonia in the polarizations, as in the cross sections)
would be a challenge to NRQCD,
where the two (necessarily different) singlet terms play a leading role.

This Letter reports the first measurement of the polarizations of promptly produced \ChicOne and \ChicTwo mesons,
using proton-proton ($\Pp\Pp$) data collected at the LHC by the CMS experiment at a center-of-mass energy of $\sqrt{s} = 8$\TeV,
corresponding to an integrated luminosity of 19.1\fbinv.
The central feature of the CMS apparatus is a superconducting solenoid of 6\unit{m} internal diameter,
providing a magnetic field of 3.8\unit{T}.
Within the solenoid volume are a silicon pixel and strip tracker, a lead tungstate crystal electromagnetic calorimeter,
and a brass and scintillator hadron calorimeter, each composed of a barrel and two endcap sections.
Muons are detected in gas-ionization chambers embedded in the steel flux-return yoke outside the solenoid.
A detailed description of the CMS detector, together with a definition of the coordinate system used and
relevant kinematic variables, can be found in Ref.~\cite{bib:CMSJINST}.

The event sample was collected with a two-level trigger system~\cite{bib:CMStrigger}.
At level-1, custom hardware processors select events with two muons.
The high-level trigger requires an opposite-sign muon pair of invariant mass 2.8--3.35\GeV,
a dimuon vertex fit $\chi^2$ probability larger than 0.5\%,
and a distance of closest approach between the two muons smaller than 0.5\unit{cm}.
The trigger also requires that the dimuon has $\pt > 7.9$\GeV and rapidity $\abs{y} < 1.25$.
The offline reconstruction requires two oppositely charged muons matching those that triggered the detector readout.
The muon tracks must pass high-purity track quality requirements~\cite{TRK-11-001},
have $\pt  > 3.5$\GeV, $\abs{\eta} < 1.6$, and
fulfill the soft muon identification requirements~\cite{bib:softmuon},
which imply, in particular, more than five hits in the silicon tracker, of which at least one is in the pixel layers.
The muons are combined to form \JPsi candidates, which are kept for further processing if $\abs{y} < 1.2$ and $8 < \pt < 30$\GeV.
Promptly produced \JPsi mesons are selected by requiring the distance between the dimuon vertex and the interaction point
be smaller than 2.5 times its uncertainty.

The analysis uses $\Chic \to \JPsi \, \gamma$ decays, with the \JPsi decaying to a dimuon.
The photons are detected through their conversions to $\Pep\Pem$ in the beam pipe and in the material of the silicon tracker,
starting from two oppositely charged tracks, of which one has at least four tracker hits and the other at least three.
The tracks must have
a conversion vertex at least 1.5\cm away from the beam axis and a $\chi^2$ probability of the kinematic fit
imposing zero mass and a common vertex that exceeds 0.05\%.
A more detailed account of the reconstruction and selection procedures is given in Refs.~\cite{bib:BPH-11-010,bib:BPH-13-005}.
The photons must have $\pt > 0.4$\GeV and $|\eta| < 1.5$.
If the distance along the beam axis between the dimuon vertex and the extrapolated photon trajectory
is smaller than 5\mm, a \Chic candidate is formed through a kinematic fit of the dimuon-photon system,
constraining the dimuon mass to the \JPsi mass~\cite{bib:PDG}, the dielectron mass to zero,
and requiring that the two muons and the photon have a common vertex.
Only \Chic candidates with a fit $\chi^2$ probability larger than 1\%
and invariant mass between 3.2 and 3.75\GeV are kept in the evaluation of the \ChicOne and \ChicTwo yields.
After all selection criteria,
around 103\,000, 106\,000, and 45\,000 \Chic candidates are kept
in the \JPsi \pt bins 8--12, 12--18, and 18--30\GeV, respectively.

The seemingly natural way to measure the \ChicOne and \ChicTwo polarizations
is to determine the angular distribution of the considered \Chic decay;
in the present case, this means the distribution of the photon direction in the \Chic rest frame.
However, that distribution depends not only on the \Chic angular momentum composition,
but also, and possibly in a very significant way,
on the (poorly known) contributions of photons with large orbital angular momentum ($J^\gamma > 1$).
A cleaner determination of the \Chic polarization is obtained by measuring the dimuon angular decay distribution
in the rest frame of the daughter \JPsi~\cite{bib:Faccioli_chic}.
It is crucial to choose as polarization axis for the \JPsi decay not the \JPsi direction in the \Chic rest frame,
as usually done in cascade decays,
but rather any axis (center-of-mass helicity or Collins--Soper~\cite{bib:CS}, for instance)
defined in terms of the beam momenta in the \JPsi rest frame
and ignoring its origin, as if it were observed inclusively.
With the latter choice, the shape of the dimuon distribution represents an exact ``clone'' of the photon distribution
in the \Chic rest frame, as it would be if it were undressed of its higher-order multipole contributions.
This method provides, therefore, a full sensitivity to the angular momentum state of the \Chic,
resulting in a (theoretically and experimentally) cleaner polarization measurement.
The present analysis is performed in the center-of-mass helicity frame~\cite{bib:Faccioli_EPJC69}
and does not use the measured photon momentum,
except to select, through the $\JPsi \, \gamma$ invariant mass distribution, the \JPsi mesons
resulting from \ChicOne or \ChicTwo decays.
The dimuon angular decay distribution is parametrized with the function
$1 + \lth \cos^2\vartheta + \lph \sin^2\vartheta \cos 2\varphi + \ltpp \sin 2\vartheta \cos\varphi$,
where $\vartheta$ and $\varphi$ are the polar and azimuthal coordinates of the
positive muon direction in the \JPsi rest frame, the system of axes
being defined with $z$ in the direction of the polarization axis and $y$
perpendicular to the production plane.
The \Chic angular momentum composition is encoded in the shape parameters \lth, \lph, and \ltpp,
whose values depend on the choice of polarization frame but
must always be within certain physical domains~\cite{bib:Faccioli_chic},
narrower than the parameter space of inclusive vector-particle production~\cite{bib:Faccioli-PRD-FrameInv,bib:Faccioli_LTVio}.
The relation between the shape parameters and the polarization configuration depends on the quarkonium state.
For example,
$\lth = +1$ indicates $J_z = \pm 1$ for the \JPsi, $J_z = 0$ for the \ChicOne, and $J_z = +2$ for the \ChicTwo;
conversely, states in the $J_z = 0$ angular momentum configuration lead to
$\lth = -1$ for the \JPsi, $\lth = +1$ for the \ChicOne, and $\lth = -0.6$ for the \ChicTwo.

The measurement of the $\lambda$ parameters implies knowing the shapes
of the \ChicOne and \ChicTwo differential cross sections as functions of \abscosth and $\varphi$,
which crucially depend on the accuracy of the corrections of the muon and photon detection efficiencies.
These efficiencies change by an order of magnitude in the low \pt bin covered by the present analysis
and shape variations within their uncertainties lead to very different \lth values.
Increasing the muon \pt threshold to avoid the turn-on region of the
efficiency function would imply a strong reduction in the number of selected events
and a smaller coverage of the \abscosth variable, effectively preventing the evaluation of \lth.
Instead, the difference between the \ChicOne and \ChicTwo polarizations,
measured from the angular dependence of the $\ChicTwo / \ChicOne$ yield ratio,
is essentially insensitive to the detection efficiencies, given that they cancel to a large extent in that ratio.

The \abscosth and $\varphi$ dependences of the yield ratio
are independently determined in three \JPsi \pt bins: 8--12, 12--18, and 18--30\GeV.
For the study of possible azimuthal dependences of the $\ChicTwo / \ChicOne$ yield ratio,
the events are split into subsamples corresponding to six equidistant $\varphi$ bins between 0 and 90$^\circ$.
Folding $\varphi$ into the first quadrant reduces the effect of the statistical fluctuations without any loss of information,
given the four-fold $\varphi$ symmetry that the angular distributions obey.
For each \pt bin, the six $\JPsi \, \gamma$ invariant mass distributions are simultaneously fitted
with an unbinned maximum likelihood fit.
In the mass fit model, identical for all $\varphi$ bins, each of the \ChicOne and \ChicTwo signal peaks
is represented by a double-sided Crystal Ball (CB) function~\cite{bib:CB},
which complements a Gaussian core distribution with lower and upper power-law tails.
The underlying combinatorial background, reflecting uncorrelated $\JPsi \, \gamma$ associations,
is parametrized by an exponential function multiplied by a term that provides a low-mass turn-down,
$(1 + \erf ((m - \mu^{\text{bg}}) \, / \, \sigma^{\text{bg}}) ) \, \exp(- m / \lambda^{\text{bg}})$,
where $m$ is the $\JPsi \, \gamma$ invariant mass
and $\mu^{\text{bg}}$, $\sigma^{\text{bg}}$, and $\lambda^{\text{bg}}$ are shape parameters.
Although the results of this analysis are insensitive to the presence of a small peak
reflecting the \ChicZero decays, the fit model includes this background term,
represented by a Breit--Wigner convolved with a Gaussian resolution function.
To minimize fit instabilities, the \ChicZero shape and yield parameters are determined from
the corresponding parameters of the \ChicOne term.
The simultaneous fit has the advantage of reducing by a factor of six the number of free parameters
defining the shapes of the signal and background mass models,
by requiring that those parameters are independent of $\varphi$,
an assumption validated by studies of simulated and measured event samples.

To study the polar angle dependence of the $\ChicTwo / \ChicOne$ yield ratio,
6, 7, or 5 \abscosth bins are considered, depending on the \pt bin.
The \abscosth coverage is smaller in the lowest \pt bin (up to 0.45 instead of up to 0.625)
because those events are the ones most affected by the single-muon \pt cut.
Analogously to the procedure just described for the $\varphi$ dimension,
the $\ChicTwo / \ChicOne$ yield ratios are obtained as a function of \abscosth
through a simultaneous fit of the $\JPsi \, \gamma$ invariant mass distributions.
In this case, however, some of the shape parameters are not required to be independent of \abscosth.
More details can be found in Ref.~\cite{bib:ThomasPhD}.

\begin{figure}[!ht]
\centering
\includegraphics[width=\cmsFigWidth]{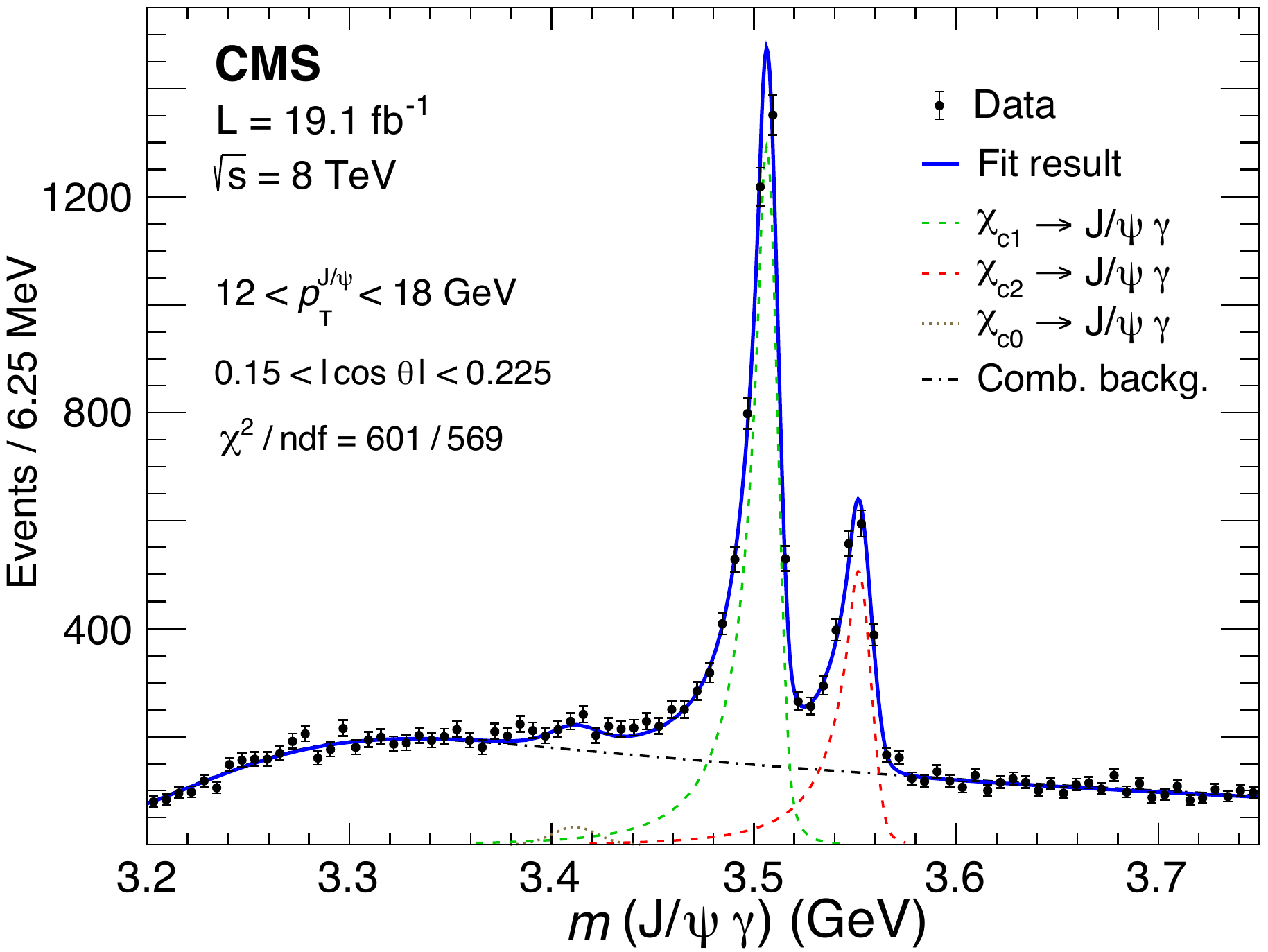}
\caption{Example of a fitted $\JPsi \, \gamma$ invariant mass distribution,
for the $0.15 < \abscosth < 0.225$ bin, in the 12--18\GeV \pt bin.
The vertical bars on the points indicate the statistical uncertainties.
The lines show the various fit contributions.}
\label{fig:mass_fit}
\end{figure}

Figure~\ref{fig:mass_fit} shows
one of the simultaneously fitted $\JPsi \, \gamma$ invariant mass distributions.
The two signal peaks are well resolved, with widths around 6\MeV,
consistent with the predictions from simulation.
All of the fitted \Chic mass distributions show good fit qualities,
as judged from the $\chi^2$ between the binned distributions and the fitted functions,
the worst case giving $\chi^2 = 601$ for 569 degrees of freedom (ndf).

For each bin in \JPsi \pt and \abscosth, or $\varphi$,
the fitted $\JPsi \, \gamma$ invariant mass distributions provide functions
reflecting the probability that an event of mass $m$ is a \ChicOne or a \ChicTwo.
The \ChicOne and \ChicTwo yields, corrected for acceptance and efficiencies,
are then computed as the sums, over all events in that
bin of \JPsi \pt and \abscosth, or $\varphi$,
of the product between the corresponding probabilities and the weights $1/\mathcal{A}_J(\abscosth, \varphi, \pt)$,
where $\mathcal{A}_J(\abscosth, \varphi, \pt)$ are the acceptance times efficiency three-dimensional maps,
independently evaluated for each \ChicJ state with large samples of simulated events.
By correcting the detector acceptance and efficiency effects on an event-by-event basis,
with weights depending on three dimuon observables (\abscosth, $\varphi$, and \pt),
this procedure is immune to integration biases affecting certain one-dimensional analyses~\cite{bib:Faccioli:2012nv}.
Simulation studies have shown that, if the three-dimensional correction maps are sufficiently fine-grained,
the results do not depend on the polarization scenario nor on the \pt distributions assumed in the simulation,
and that all physically allowed differences between the \ChicOne and \ChicTwo polarizations, in any frame,
can be reliably determined from the dependences of the $\ChicTwo / \ChicOne$ yield ratios on \abscosth and $\varphi$.

\begin{figure}[!th]
\centering
\includegraphics[width=\cmsFigWide]{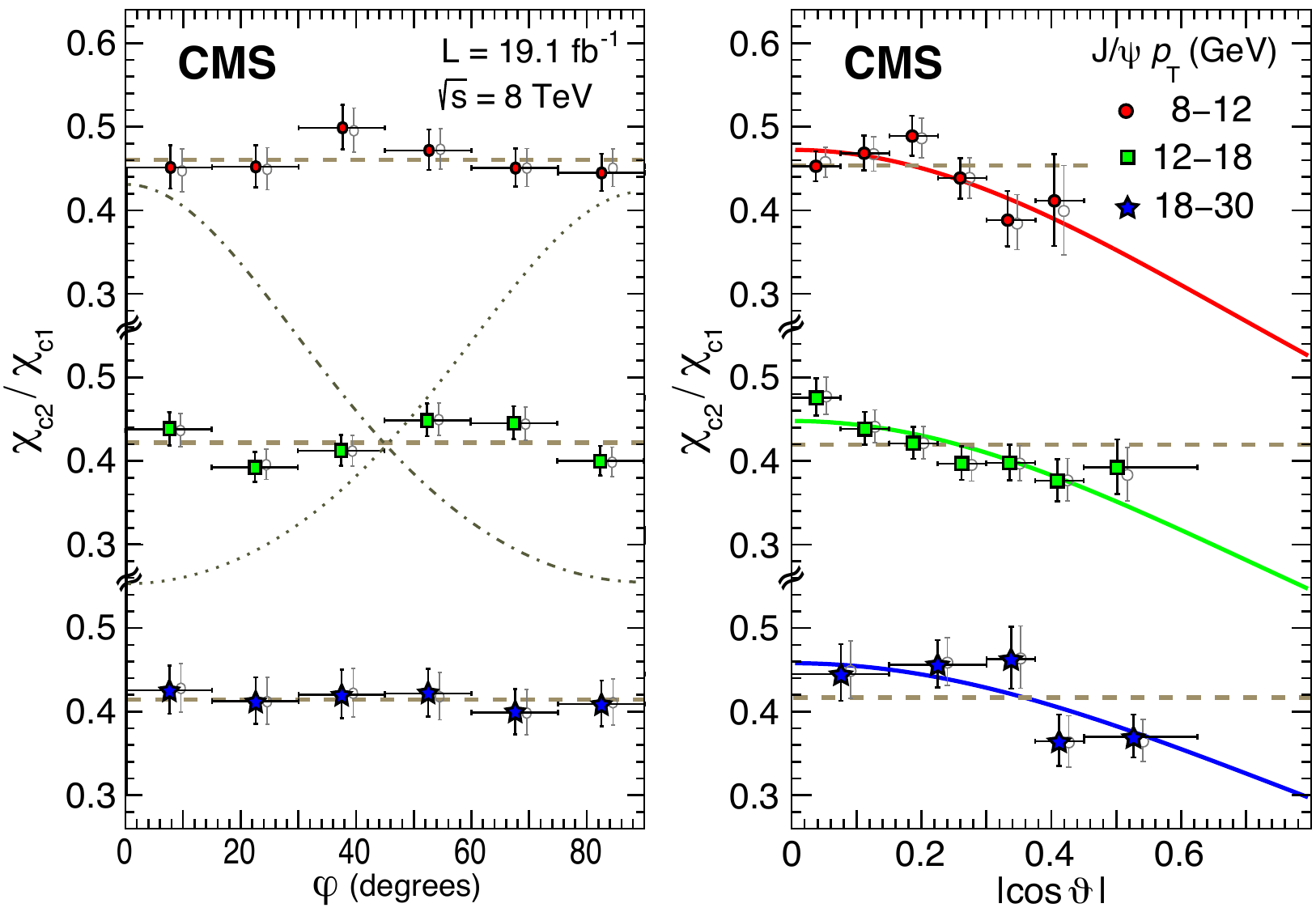}
\caption{The $\ChicTwo / \ChicOne$ yield ratio vs.\ $\varphi$ (left) and \abscosth (right),
for the three \JPsi \pt bins.
The grey markers (slightly shifted horizontally)
show the values before acceptance and efficiency corrections,
scaled vertically for an easier shape comparison.
The vertical bars represent the statistical uncertainties and the horizontal bars the bin widths.
The solid and dashed curves show, respectively, the NRQCD~\cite{bib:Faccioli:EPJC78p268}
and unpolarized scenarios.
The dotted and dash-dotted curves
illustrate maximally different natural polarizations in the Collins--Soper frame,
leading to large differences in azimuthal anisotropy.}
\label{fig:chictwo_over_chicone}
\end{figure}

The corrected ratios are reported in
Tables~\suppMaterialRef{1}{\ref{tab:ratio_phi}} and~\suppMaterialRef{2}{\ref{tab:ratio_costh}} of \suppMaterial,
and shown in Fig.~\ref{fig:chictwo_over_chicone},
where it can be seen that the uncorrected and corrected values are almost identical,
apart from normalization factors irrelevant for the determination of the polar and azimuthal anisotropies.

Several sources of potential systematic effects have been considered,
by redoing the analysis with different inputs
and comparing the obtained results with the nominal ones.
The results are insensitive to variations of the
thresholds used to reject the nonprompt contamination from \cPqb hadron decays,
estimated to be around 5\%,
or events with a poor kinematic vertex fit quality in the reconstruction of the \Chic candidates.
The fits of the mass distributions were redone
using alternative options for the low- and high-mass tails of the double-sided CB functions,
and by varying the combinatorial background description, both by changing the floating parameters
of the nominal function and by using the alternative function $(x - x_0)^\lambda \, \exp{(\nu(x - x_0))}$,
where $\nu$ is left free, $\lambda$ is fitted to a constant, and $x_{0} = 3.2$\GeV,
a value determined in fits to the background-only mass distributions
obtained by excluding the 3.37--3.6\GeV region.
The sensitivity of the results to the acceptance and efficiency corrections was evaluated by
redoing the analysis with maps computed with alternative single-muon and photon detection
efficiencies, as well as with
simulated samples generated with different $\pt/M$ shapes for each of the two \Chic states.
All effects lead to similar variations in the yields of the two states and cancel,
to a large extent, in the $\ChicTwo / \ChicOne$ ratio,
apart from a normalization shift that has no impact on the angular anisotropies.
The total systematic uncertainties are less than 20\% of the statistical ones.

The \ChicTwo to \ChicOne yield ratios as a function of $\varphi$,
shown in Fig.~\ref{fig:chictwo_over_chicone}~(left),
are compatible with being flat, excluding large differences in azimuthal anisotropy,
as exemplified by the two curves compared to the data points in the second \pt bin.
These curves represent the simplest conceivable polarization hypotheses
leading to large azimuthal effects in the helicity frame:
\ChicOne and \ChicTwo have maximally different polar anisotropies in the
Collins--Soper frame, corresponding to specific alignments of their angular momentum vectors
along the collision direction
($J_{z}^{\ChicOne} = J_{z}^{\ChicTwo} = 0$ and
$J_{z}^{\ChicOne} = \pm 1$, $J_{z}^{\ChicTwo} = \pm 2$,
for the dotted and dash-dotted curve, respectively).
In fact, the change from the Collins--Soper to the helicity quantization axis is almost a 90$^\circ$
rotation, transforming polarized distributions into azimuthally anisotropic ones.
This uniform $\varphi$ behavior confirms the choice of the helicity axis as the one that
should reflect most closely the natural alignment of the angular momentum vector,
maximizing the polar anisotropy effects.

In Fig.~\ref{fig:chictwo_over_chicone}~(right)
the measured \abscosth dependence of the $\ChicTwo / \ChicOne$ ratio
is compared to the analytic expression
$(1 + \lth^{\ChicTwo}\cos^{2}\vartheta) \, / \, (1 + \lth^{\ChicOne}\cos^{2}\vartheta)$.
Two scenarios are considered.
The unpolarized scenario, $\lth^{\ChicOne} = \lth^{\ChicTwo} = 0$ independently of \pt,
represented in Fig.~\ref{fig:chictwo_over_chicone}~(right) by the dashed flat lines,
gives a poor description of the data.
A fit with free normalizations leads to $\chi^2/\textrm{ndf} = 31/15$,
corresponding to a $\chi^2$ probability of 0.9\%.
The NRQCD scenario~\cite{bib:Faccioli:EPJC78p268},
where $\lth^{\ChicOne} = 0.72$, 0.65, and 0.56,
and $\lth^{\ChicTwo} = -0.48$, $-0.35$, and $-0.19$,
for the average \pt values in each of the three bins,
agrees well with the data: $\chi^2/\textrm{ndf} = 13/15$,
corresponding to $P(\chi^2) = 58$\%.

\begin{figure}[!t]
\centering
\includegraphics[width=0.95\cmsFigWidth]{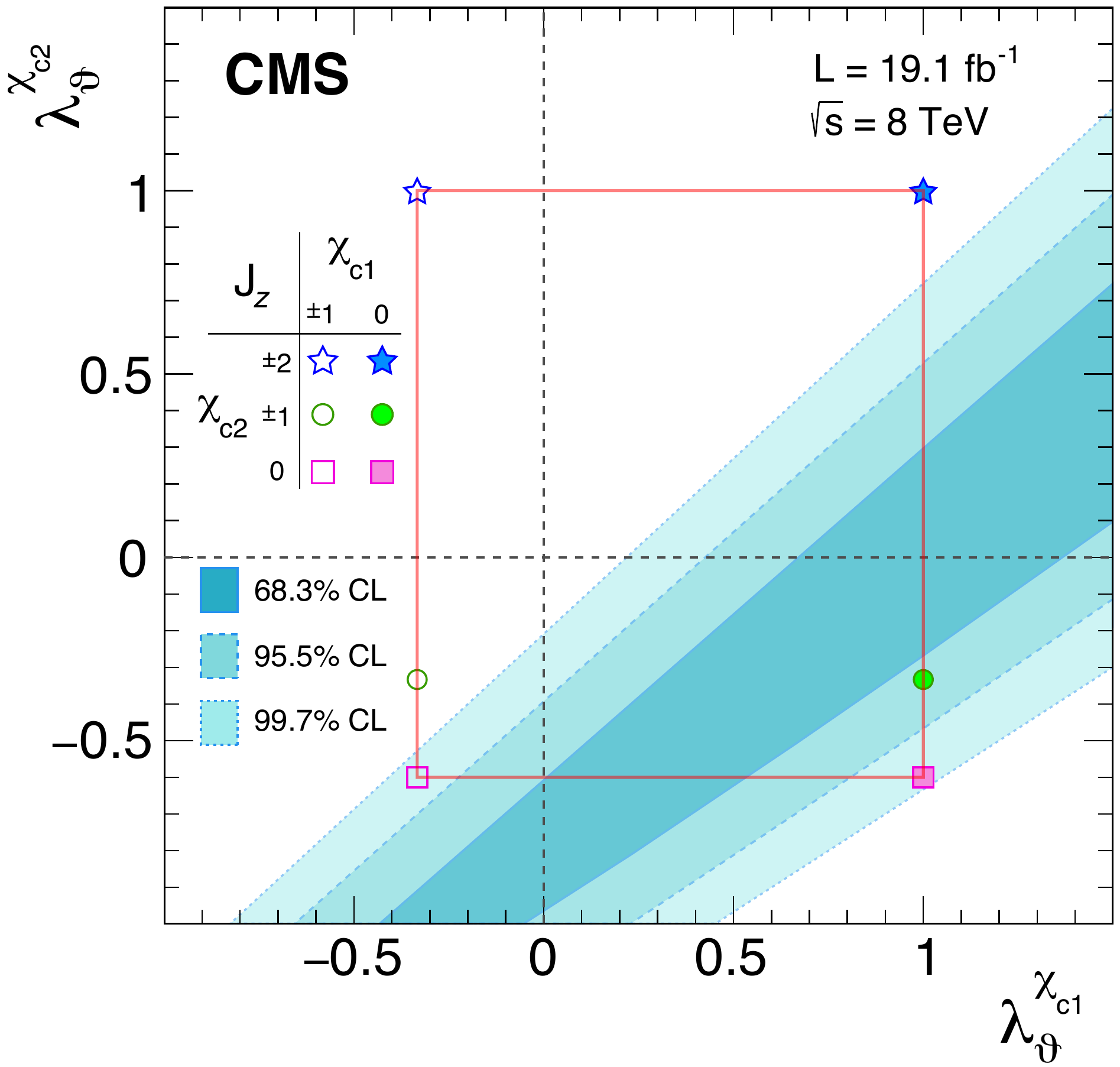}
\caption{Two-dimensional $\lth^{\ChicTwo}$ vs.\ $\lth^{\ChicOne}$ contours, at 68.3, 95.5, and 99.7\% confidence levels (CL),
measured combining the three \JPsi \pt bins. The physically allowed region (red rectangle) and
six pure angular momentum configurations (markers) are also shown.
The crossing of the two dashed lines represents the unpolarized case.}
\label{fig:2dplot}
\end{figure}

Figure~\ref{fig:2dplot} shows the polar anisotropy parameters $\lth^{\ChicOne}$ and $\lth^{\ChicTwo}$
derived from the measured \abscosth dependence of the $\ChicTwo / \ChicOne$ ratio,
combining the three \pt bins.
The contours in the $\lth^{\ChicOne}$ vs.\ $\lth^{\ChicTwo}$ plane are obtained by scanning
the two $\lth$ parameters and the three normalizations
to evaluate the $\chi^2$ profiles corresponding to the 68.3, 95.5, and 99.7\% confidence levels.
The unpolarized scenario ($\lth^{\ChicOne} = \lth^{\ChicTwo} = 0$),
as well as more than half of the physically allowed region,
including all cases where $\lth^{\ChicTwo} \ge \lth^{\ChicOne}$,
are outside the 99.7\% contour.
In terms of specific pure angular momentum configurations,
it can be seen that, in particular, the cases
$J_{z}^{\ChicTwo} = \pm 2$ and
$J_{z}^{\ChicOne} = J_{z}^{\ChicTwo} = \pm 1$ are strongly disfavored.

The correlation between the $\lth^{\ChicOne}$ and $\lth^{\ChicTwo}$ parameters
can be accurately expressed through a simple parametrization:
$\lth^{\ChicTwo} = (-0.94 + 0.90 \, \lth^{\ChicOne}) \pm (0.51 + 0.05 \, \lth^{\ChicOne})$,
$(-0.76 + 0.80 \, \lth^{\ChicOne}) \pm (0.26 + 0.05 \, \lth^{\ChicOne})$, and
$(-0.78 + 0.77 \, \lth^{\ChicOne}) \pm (0.26 + 0.06 \, \lth^{\ChicOne})$,
for the three consecutive \pt bins.
These expressions can be used for direct comparisons to theoretical scenarios.

\begin{figure}[t]
\centering
\includegraphics[width=\cmsFigWide]{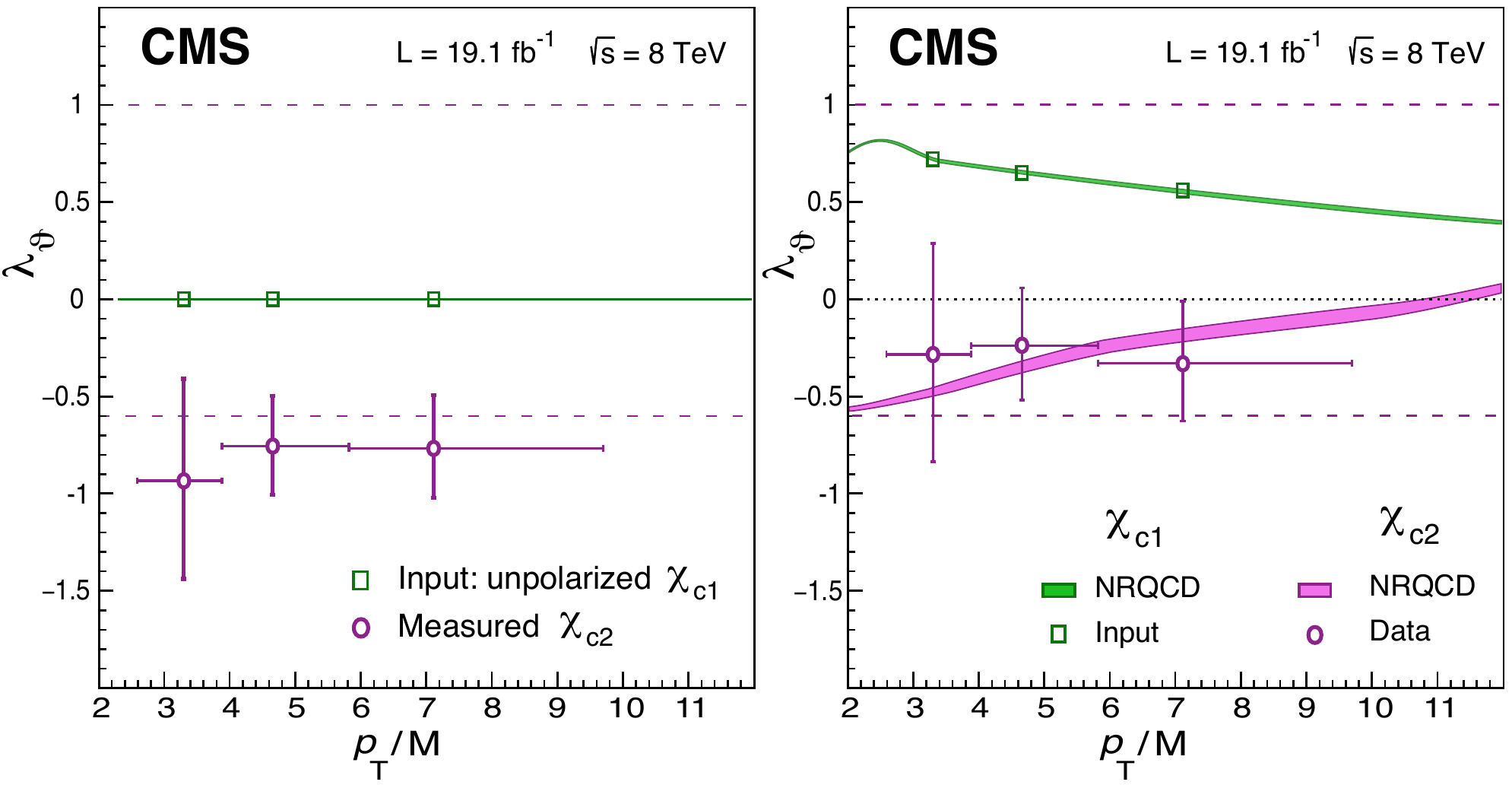}
\caption{The $\lth^{\ChicTwo}$ values (circles) measured
when the $\lth^{\ChicOne}$ values (squares) are fixed to the unpolarized (left)
or the NRQCD (right) scenarios (green curves), as a function of $\pt/M$ of the \JPsi.
The purple band on the right is the NRQCD prediction for $\lth^{\ChicTwo}$~\cite{bib:Faccioli:EPJC78p268},
while in the unpolarized scenario $\lth^{\ChicTwo} = \lth^{\ChicOne} = 0$.
The markers are shown at the average $\pt/M$ values in each bin,
the vertical bars represent the total uncertainties, and the horizontal bars the bin widths.
The dashed lines indicate the physically allowed range of $\lth^{\ChicTwo}$.}
\label{fig:pt_dependences}
\end{figure}

Figure~\ref{fig:pt_dependences} shows, as a function of $\pt/M$ of the \JPsi
(equal on average to the $\pt/M$ of the \ChicOne and \ChicTwo mothers~\cite{bib:Faccioli:PLB773}),
the $\lth^{\ChicTwo}$ values measured when $\lth^{\ChicOne}$ is fixed to the
predictions of the two scenarios already considered in Fig.~\ref{fig:chictwo_over_chicone}.
Setting $\lth^{\ChicOne} = 0$ leads to $\lth^{\ChicTwo}$ values that are
significantly different from zero.
The NRQCD prediction is, instead, in good agreement with the measurement.

In summary, first experimental constraints
on the polarizations of promptly produced \ChicOne and \ChicTwo mesons
have been obtained, using $\Pp\Pp$ collisions at $\sqrt{s} = 8\TeV$.
The analysis uses the $\JPsi \, \gamma$ decay channel in three \JPsi \pt bins between 8 and 30\GeV.
The measurement, made in the helicity frame,
shows a significant difference between the polar anisotropy parameters $\lth^{\ChicOne}$ and $\lth^{\ChicTwo}$,
in agreement with the NRQCD prediction.
This result is a new step in the experimental studies of quarkonium production and the
first significant indication of kinematic differences between the various quarkonia.

\begin{acknowledgments}
We congratulate our colleagues in the CERN accelerator departments for the excellent performance of the LHC and thank the technical and administrative staffs at CERN and at other CMS institutes for their contributions to the success of the CMS effort. In addition, we gratefully acknowledge the computing centers and personnel of the Worldwide LHC Computing Grid for delivering so effectively the computing infrastructure essential to our analyses. Finally, we acknowledge the enduring support for the construction and operation of the LHC and the CMS detector provided by the following funding agencies: BMBWF and FWF (Austria); FNRS and FWO (Belgium); CNPq, CAPES, FAPERJ, FAPERGS, and FAPESP (Brazil); MES (Bulgaria); CERN; CAS, MoST, and NSFC (China); COLCIENCIAS (Colombia); MSES and CSF (Croatia); RPF (Cyprus); SENESCYT (Ecuador); MoER, ERC IUT, PUT and ERDF (Estonia); Academy of Finland, MEC, and HIP (Finland); CEA and CNRS/IN2P3 (France); BMBF, DFG, and HGF (Germany); GSRT (Greece); NKFIA (Hungary); DAE and DST (India); IPM (Iran); SFI (Ireland); INFN (Italy); MSIP and NRF (Republic of Korea); MES (Latvia); LAS (Lithuania); MOE and UM (Malaysia); BUAP, CINVESTAV, CONACYT, LNS, SEP, and UASLP-FAI (Mexico); MOS (Montenegro); MBIE (New Zealand); PAEC (Pakistan); MSHE and NSC (Poland); FCT (Portugal); JINR (Dubna); MON, RosAtom, RAS, RFBR, and NRC KI (Russia); MESTD (Serbia); SEIDI, CPAN, PCTI, and FEDER (Spain); MOSTR (Sri Lanka); Swiss Funding Agencies (Switzerland); MST (Taipei); ThEPCenter, IPST, STAR, and NSTDA (Thailand); TUBITAK and TAEK (Turkey); NASU (Ukraine); STFC (United Kingdom); DOE and NSF (USA).
\end{acknowledgments}

\bibliography{auto_generated}

\ifthenelse{\boolean{cms@external}}{}{
\clearpage
\appendix
\numberwithin{table}{section}
\section{Numerical values of the measured yield ratios}
\label{app:tables}
\begin{table}[h!]
\centering
\topcaption{The ratio of the \ChicTwo to \ChicOne yields, corrected for acceptance and efficiencies,
vs.\ $\varphi$, in three \JPsi \pt ranges. The average $\varphi$ values are also given.}
\label{tab:ratio_phi}
\renewcommand{\arraystretch}{1.5}
\begin{scotch}{c c c c}
\JPsi \pt ({\GeVns})
& $\varphi$ (degrees) & $\langle \varphi \rangle$ (degrees) & $\ChicTwo / \ChicOne$ \\
\hline
\multirow{6}{*}{8--12}
&  ~~0--15 &  ~~7.8 & $0.451^{+0.027}_{-0.025}$ \\
& 15--30 & 22.6 & $0.452^{+0.026}_{-0.025}$ \\
& 30--45 & 37.6 & $0.499^{+0.027}_{-0.026}$ \\
& 45--60 & 52.6 & $0.472^{+0.025}_{-0.024}$ \\
& 60--75 & 67.6 & $0.450^{+0.023}_{-0.022}$ \\
& 75--90 & 82.5 & $0.445^{+0.023}_{-0.022}$ \\
\hline
\multirow{6}{*}{12--18}
&  ~~0--15 &  ~~7.7 & $0.438^{+0.021}_{-0.020}$ \\
& 15--30 & 22.5 & $0.393^{+0.018}_{-0.017}$ \\
& 30--45 & 37.5 & $0.412^{+0.019}_{-0.018}$ \\
& 45--60 & 52.4 & $0.449^{+0.020}_{-0.019}$ \\
& 60--75 & 67.5 & $0.445^{+0.020}_{-0.019}$ \\
& 75--90 & 82.5 & $0.400^{+0.018}_{-0.017}$ \\
\hline
\multirow{6}{*}{18--30}
&  ~~0--15 &  ~~7.6 & $0.425^{+0.030}_{-0.028}$ \\
& 15--30 & 22.6 & $0.412^{+0.028}_{-0.027}$ \\
& 30--45 & 37.5 & $0.420^{+0.030}_{-0.028}$ \\
& 45--60 & 52.5 & $0.421^{+0.030}_{-0.028}$ \\
& 60--75 & 67.6 & $0.399^{+0.028}_{-0.026}$ \\
& 75--90 & 82.5 & $0.409^{+0.028}_{-0.027}$ \\
\end{scotch}
\end{table}

\begin{table}[h!]
\centering
\topcaption{The ratio of the \ChicTwo to \ChicOne yields, corrected for acceptance and efficiencies,
vs.\ \abscosth, in three \JPsi \pt ranges. The average \abscosth values are also given.
Fitting these ratios to a flat function (unpolarized scenario) leads to $\chi^2/\textrm{ndf} = 7.2/5$, $13.5/6$, and $10.3/4$, 
respectively for the \pt ranges 8--12, 12--18, and 18--30\GeV; 
the corresponding values for the NRQCD prediction are $4.1/5$, $4.9/6$, and $4.2/4$.}
\label{tab:ratio_costh}
\renewcommand{\arraystretch}{1.5}
\begin{scotch}{c c c c}
\JPsi \pt ({\GeVns})
& \abscosth & $\langle \abscosth \rangle$ & $\ChicTwo / \ChicOne$ \\
\hline
\multirow{6}{*}{8--12}
& 0.000--0.075 & 0.037 & $0.453^{+0.018}_{-0.018}$ \\
& 0.075--0.150 & 0.111 & $0.468^{+0.021}_{-0.020}$ \\
& 0.150--0.225 & 0.185 & $0.489^{+0.025}_{-0.024}$ \\
& 0.225--0.300 & 0.259 & $0.439^{+0.024}_{-0.025}$ \\
& 0.300--0.375 & 0.332 & $0.388^{+0.035}_{-0.031}$ \\
& 0.375--0.450 & 0.404 & $0.411^{+0.056}_{-0.054}$ \\
\hline
\multirow{7}{*}{12--18}
& 0.000--0.075 & 0.038 & $0.476^{+0.023}_{-0.021}$ \\
& 0.075--0.150 & 0.113 & $0.438^{+0.020}_{-0.019}$ \\
& 0.150--0.225 & 0.187 & $0.421^{+0.020}_{-0.019}$ \\
& 0.225--0.300 & 0.262 & $0.397^{+0.021}_{-0.019}$ \\
& 0.300--0.375 & 0.336 & $0.398^{+0.022}_{-0.021}$ \\
& 0.375--0.450 & 0.409 & $0.376^{+0.026}_{-0.024}$ \\
& 0.450--0.625 & 0.502 & $0.392^{+0.033}_{-0.032}$ \\
\hline
\multirow{5}{*}{18--30}
& 0.000--0.150 & 0.076 & $0.445^{+0.036}_{-0.032}$ \\
& 0.150--0.300 & 0.225 & $0.456^{+0.030}_{-0.027}$ \\
& 0.300--0.375 & 0.338 & $0.463^{+0.039}_{-0.036}$ \\
& 0.375--0.450 & 0.412 & $0.365^{+0.032}_{-0.030}$ \\
& 0.450--0.625 & 0.526 & $0.370^{+0.027}_{-0.025}$ \\
\end{scotch}
\end{table}

}
\cleardoublepage \section{The CMS Collaboration \label{app:collab}}\begin{sloppypar}\hyphenpenalty=5000\widowpenalty=500\clubpenalty=5000\vskip\cmsinstskip
\textbf{Yerevan Physics Institute, Yerevan, Armenia}\\*[0pt]
A.M.~Sirunyan$^{\textrm{\dag}}$, A.~Tumasyan
\vskip\cmsinstskip
\textbf{Institut f\"{u}r Hochenergiephysik, Wien, Austria}\\*[0pt]
W.~Adam, F.~Ambrogi, T.~Bergauer, M.~Dragicevic, J.~Er\"{o}, A.~Escalante~Del~Valle, M.~Flechl, R.~Fr\"{u}hwirth\cmsAuthorMark{1}, M.~Jeitler\cmsAuthorMark{1}, N.~Krammer, I.~Kr\"{a}tschmer, D.~Liko, T.~Madlener, I.~Mikulec, N.~Rad, J.~Schieck\cmsAuthorMark{1}, R.~Sch\"{o}fbeck, M.~Spanring, W.~Waltenberger, C.-E.~Wulz\cmsAuthorMark{1}, M.~Zarucki
% ...
\vskip\cmsinstskip
\dag: Deceased\\
1:  Also at Vienna University of Technology, Vienna, Austria\\
% ...
\end{sloppypar}
\end{document}